\documentclass{aa}
\usepackage{graphicx}
\usepackage{url}
\usepackage{natbib}
\usepackage{txfonts}
   \title{Photometric redshifts with Quasi Newton Algorithm (MLPQNA). Results in the PHAT1 contest.}

\begin{document}

   \author{S. Cavuoti
          \inst{1}\fnmsep\inst{2}
          \and
          M. Brescia\inst{2}\fnmsep\inst{1}
          \and
          G. Longo\inst{1}\fnmsep\inst{2}\fnmsep\inst{3}
          \and
          A. Mercurio\inst{2}
          %\thanks{Just to show the usage of the elements in the author field}
          }

   \institute{Department of Physics, Federico II University, via Cinthia 6, I-80126 Napoli, Italy
   					 \email{cavuoti@na.infn.it}
         		 \and
             INAF - Astronomical Observatory of Capodimonte, via Moiariello 16, I-80131 Napoli, Italy              		
             \and
             Visiting associate - Department of Astronomy, California Institute of Technology, CA 90125, USA
             %\thanks{The university of heaven temporarily does not accept e-mails}
             }

   \date{Received June 2012; accepted August 2012}

% \abstract{}{}{}{}{}
% 5 {} token are mandatory

  \abstract
  % context heading (optional)
  % {} leave it empty if necessary
  {Since the advent of modern multiband digital sky surveys, photometric redshifts (photo-z's) have become relevant if not crucial
   to many fields of observational cosmology, from the characterization of cosmic structures, to weak and strong lensing.}
  % aims heading (mandatory)
   {We describe an application to an astrophysical context, namely the evaluation of photometric redshifts, of MLPQNA, a machine learning  method based on Quasi Newton Algorithm.}
  % methods heading (mandatory)
   {Theoretical methods for photo-z's evaluation are based on the interpolation of a priori knowledge (spectroscopic redshifts or SED templates) and represent an ideal comparison ground for neural networks based methods. The MultiLayer Perceptron with Quasi Newton learning rule (MLPQNA) described here is a computing effective implementation of Neural Networks for the first time exploited to solve regression problems in the astrophysical context and is offered to the community through the DAMEWARE (DAta Mining \& Exploration Web Application REsource) infrastructure. }
  % results heading (mandatory)
   {The PHAT contest (\citealt{hildebrandt2010}) provides a standard dataset to test old and new methods for photometric redshift evaluation and with a set of statistical indicators which allow a straightforward comparison among different methods. The MLPQNA model has been applied on the whole PHAT1 dataset of 1984 objects after an optimization of the model performed by using as training set the 515 available spectroscopic redshifts. When applied to the PHAT1 dataset, MLPQNA obtains the best bias accuracy (0.0006) and very competitive accuracies in terms of scatter (0.056) and outlier percentage (16.3\%), scoring as the second most effective empirical method among those which have so far participated to the contest. MLPQNA shows better generalization capabilities than most other empirical methods especially in presence of underpopulated regions of the Knowledge Base.}
  % conclusions heading (optional), leave it empty if necessary
   {}


\authorrunning{Cavuoti et al. 2012}
\titlerunning{MLPQNA: A Quasi Newton based Neural Model for Photometric Redshifts}

  \maketitle
%
%________________________________________________________________

\section{Introduction}

Estimating redshifts of celestial objects is one of the most
  pressing technological issues in the observational astronomy and, since
  the advent of modern multiband digital sky surveys, photometric
  redshifts (photo-z's) have become fundamental when it is necessary
  to know the distances of million of objects over large cosmological
  volumes. Photo-z's provide redshift estimates for objects fainter
than the spectroscopic limit, and result much more efficient in terms
of the number of objects per telescope time with respect to spectroscopic ones (spec-z).
For these reasons, after the advent of modern panchromatic digital surveys, photo-z's have
become crucial. For instance, they are essential in constraining dark
matter and dark energy studies by means of weak gravitational lensing,
for the identification of galaxy clusters and groups
(e.g. \citealt{capozzi2009}), for type Ia supernovae, and to study the
mass function of galaxy clusters
\citep{albrecht2006,peacock2006,keiichi2012}. The need for fast and
reliable methods for photo-z evaluation will become even greater in
the near future for the exploitation of ongoing and planned surveys.
In fact, future large field public imaging projects, like KiDS
  (Kilo-Degree
  Survey\footnote{http://www.astro-wise.org/projects/KIDS/}), DES
  (Dark Energy Survey\footnote{http://www.darkenergysurvey.org/}),
  LSST (Large Synoptic Survey
  Telescope\footnote{http://www.lsst.org/lsst/}), and Euclid
  (\citealt{euclid}), require extremely accurate photo-z's to obtain
  accurate measurements that does not compromise the surveys
  scientific goals. This explains the very rapid growth in the
number of methods which can be more or less effectively used to derive
photo-z's estimates, and the efforts made to better understand and
characterize their biases and systematics.  The possibility to achieve
a very low level of residual systematics \citep{huterer2006,dabrusco2007}, is in fact strongly
influenced by many factors: the observing strategy, the accuracy of
the photometric calibration, the different point-spread-function in
different bands, the adopted de-reddening procedures, etc.  The
evaluation of photo-z's is made possible by the existence of a rather
complex correlation existing between the fluxes as measured in broad
band photometry, the morphological types of the galaxies and their
distance. The search for such correlation (a non-linear mapping
between the photometric parameter space and the redshift values) is
particularly suited for data mining methods.  Existing methods can be
broadly divided into two large groups: theoretical and empirical
methods. Theoretical methods use templates, like libraries of either observed
  galaxy spectra or model Spectral Energy Distributions (SEDs). These templates can be
  shifted to any redshift and then convolved with the transmission
  curves of the filters used in the photometric survey to create the
  template set for the redshift estimators (e.g. \citealt{koo1999}, \citealt{massarotti2001a}, \citealt{massarotti2001b}, \citealt{csabai2003}). However, for datasets in which accurate and multiband photometry for a large number of objects are complemented by  spectroscopic redshifts for a statistically significant subsample of the same objects, the empirical methods offer greater accuracy, as well as being far more efficient. These methods use the
  subsample of the photometric survey with spectroscopically-measured
  redshifts as a \textit{training set} to constrain the fit of a polynomial
function mapping the photometric data as redshift estimators.
  %This subsample describes empirically the redshift distribution in magnitude and color space and it is used to calibrate this relation. Both methods make use of training sets as bases for the redshift estimating routines.

Several template based methods have been developed
to derive photometric redshifts with increasingly high precision
such as \textit{BPZ\footnote{http://acs.pha.jhu.edu/~txitxo/bpzdoc.html}, HyperZ\footnote{http://webast.ast.obs-mip.fr/hyperz/},
Kcorrect\footnote{http://cosmo.nyu.edu/blanton/kcorrect/}, Le PHARE\footnote{http://www.cfht.hawaii.edu/~arnouts/LEPHARE/lephare.html}, ZEBRA\footnote{http://www.exp-astro.phys.ethz.ch/ZEBRA},
LRT Libraries\footnote{http://www.astronomy.ohio-state.edu/~rjassef/lrt/},  EAzY\footnote{http://www.astro.yale.edu/eazy/}, Z-PEG\footnote{http://imacdlb.iap.fr:8080/cgi-bin/zpeg/zpeg.pl}. Moreover there are also training set based methods, such as AnnZ\footnote{http://www.homepages.ucl.ac.uk/~ucapola/annz.html}, RFPhotoZ\footnote{http://www.sdss.jhu.edu/~carliles/photoZ/RFPhotoZ/} among others).}
The variety of methods and approaches and their application to
different types of datasets, as well as the adoption of different and
often not comparable statistical indicators, make it difficult to
evaluate and compare performances in an unambiguous and homogeneous
way.  Useful but limited in scope blind tests of photo-z's have been
performed in \cite{hogg1998} on spectroscopic data from the Keck
telescope on the Hubble Deep Field (HDF), in \cite{hildebrandt2008}
on spectroscopic data from the VIMOS VLT Deep Survey (VVDS;
\citealt{lefevre2004}) and the FORS Deep Field (FDF; \citealt{noll2004},
and in \citealt{abdalla2008}) on the sample of Luminous Red Galaxies
from the SDSS-DR6.

A significant advance in comparing different methods was introduced by
Hildebrandt and collaborators (\citealt{hildebrandt2010}), with the so
called PHAT (PHoto-z Accuracy Testing) contest, which adopts a
black-box approach which is typical of benchmarking. Instead of
insisting on the subtleties of the data structure, they performed a
homogeneous comparison of the performances concentrating the analysis
on the last link in the chain: the photo-z's methods themselves.

As pointed out by the authors, in fact, ''{\it it is clear that the two regimes - data and method - cannot be separated cleanly
because there are connections between the two. For example, it is highly likely that one method of photo-z estimation will
perform better than a second method on one particular dataset while the situation may well be reversed on a different data
set.}'' (cf. \citealt{hildebrandt2010}).\\

Considering that empirical methods are trained on real data and do not
require assumptions on the physics of the formation and evolution of stellar populations, Neural Networks (hereafter NNs) are
excellent tools to interpolate data and to extract patterns and trends (cf. the standard textbook by \citealt{bishop2006}). In
this paper we show the application in the PHAT1 contest of the Multi Layer Perceptron (MLP) implemented with a Quasi
Newton Algorithm (QNA) as learning rule which has been employed for the first time to interpolate the photometric redshifts.

The present work follows the same path, by having as its aim
the testing and probing of the accuracy of the Quasi Newton based Neural
Model (MLPQNA) for the derivation of photometric redshifts.
The application of MLPQNA to the photometric redshift estimation of
QSO will be presented in Brescia et al. (in preparation).

In Sect. \ref{thedata} we shortly describe the PHAT contest and the
PHAT1 data made available to the contestants and used for the present
work.  In Sect. \ref{mlpqna} we describe the MLPQNA method which was
implemented by us and used for the contest, while in Sect.
\ref{experiments} we describe the experiments performed and, in Sect.
\ref{PR} we present the results derived for us by the PHAT
board. Summary and conclusions are wrapped up in Sect. \ref{discus}.

\section{The PHAT dataset}\label{thedata}
First results from the PHAT contest were presented in
\cite{hildebrandt2010}, but the contest still continues at the
project's web site. PHAT provides a standardized test environment
which consists of simulated and observed photometric catalogues complemented
with additional materials like filter curves
convolved with transmission curves, SED templates, and training
sets. The PHAT project has been conceived as a blind contest, still open to host new participants who want to test their own regression method performances, as it was in our case, since we developed our model in the last two years.
However, the subsets used to evaluate the performances are
still kept secret in order to provide a more reliable comparison of
the various methods.  Two different datasets are available
(see \citealt{hildebrandt2010} for more details).

The first one, indicated as PHAT0, is based on a very limited
template set and a long wavelength baseline (from UV to
mid-IR). It is composed by a noise-free catalogue with accurate
synthetic colors and a catalogue with a low level of additional
noise.
PHAT0 represents an easy case to test the most basic elements
of photo-z estimation and to identify possible low-level discrepancies
between the methods.

The second one, which is the one used in the present work, is
the PHAT1 dataset, which is based on real data originating from
the Great Observatories Origins Deep Survey Northern field (GOODS-North;
\citealt{giavalisco2004}). According to \cite{hildebrandt2010}, it
represents a much more complex environment to test methods to
estimate photo-z's, pushing codes to their limits and revealing
more systematic difficulties.  Both PHAT test datasets are made
publicly available through the PHAT
website\footnote{\url{http://www.astro.caltech.edu/twiki_phat/bin/view/Main/GoodsNorth}}
while in \cite{hildebrandt2010} there is a detailed description of
the statistical indicators which were used for the comparison of the
results provided by the 21 participants who have so far participated by
submitting results obtained with 17 different photo-z codes.

The PHAT1 dataset consists of photometric observations, both from
ground and space instruments, presented in \cite{giavalisco2004},
complemented with additional data in other bands derived from
\cite{capak2004}. The final dataset covers the full UV-IR range and
includes 18 bands: U (from KPNO), B, V, R, I, Z (from SUBARU),
F435W, F606W, F775W, F850LP (from HST-ACS), J, H (from ULBCAM), HK
(from QUIRC), K (from WIRC) and 3.6, 4.5, 5.8 and 8.0 $\mu$ (from IRAC
Spitzer).

The photometric dataset was then cross correlated with spectroscopic
data from \cite{cowie2004, wirth2004, treu2005}, and \cite{reddy2006}.
Therefore, the final PHAT1 dataset consists of 1984 objects with
18-band photometry and accurate spectroscopic redshifts.  In the
publicly available dataset a little more than one quarter of the
objects comes with spectroscopic redshifts and can be used as
Knowledge Base (KB) for training empirical methods.\\
In this contest, in fact, only 515 objects were made available with the corresponding spectroscopic redshift, while for the remaining 1469 objects the related spectroscopic redshift has been hidden to all participants. The immediate consequence is that any empirical method exploited in the contest was
constrained to use the 515 objects as training set (knowledge base) and the 1469 objects as the test set, to be delivered to PHAT contest board in order to obtain back the statistical evaluation results.
While it is clear that the limited amount of objects in the knowledge base is not
sufficient to ensure the best performances of most empirical methods,
the fact that all methods must cope with similar difficulties makes
the comparison consistent.
\section{The MLPQNA regression model}\label{mlpqna}
MLPQNA stands for the traditional neural network model named Multi
Layer Perceptron (MLP; cf. \citealt{bishop2006}) implemented with a Quasi
Newton Algorithm (QNA) as learning rule.  This particular
implementation of the traditional MLP's has already been described in
\cite{brescia2012}, and we refer to that paper for a more detailed
description in the classification problem context.  MLPQNA is made available to the community through the
DAMEWARE (DAta Mining \& Exploration Web Application
REsource;
\citealt{brescia2009, brescia2011, brescia2012, brescia2012b}).  In the text we also provide the details and the parameters
settings for the best performing MLPQNA model so that anyone can
easily reproduce the results using the Web Application. User's manuals
are available on the DAMEWARE web site\footnote{\url{http://dame.dsf.unina.it/beta_info.html}}.
A complete mathematical description of the MLPQNA model is available on the DAME web site\footnote{\url{http://dame.dsf.unina.it/machine_learning.html#mlpqna}}.
Feed-forward neural networks provide a general framework for representing nonlinear functional mappings between a set of input variables and a set of output variables \citep{bishop2006}. One can achieve this goal by representing the nonlinear function of many variables by a composition of non-linear activation functions of one variable, which formally describes the mathematical representation of a feed-forward neural network with two computational layers (Eq.~\ref{PHAT:eq1}):

\begin{equation}
y_k = \sum^M_{j=0} w^{(2)}_{kj}g\left(\sum_{i=0}^d w_{ji}^{(1)}x_i \right)
\label{PHAT:eq1}
\end{equation}

\begin{figure*}
   \centering
   \includegraphics[width=7cm]{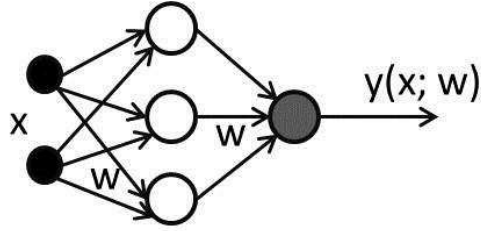}
   \caption{The classical feed-forward architecture of a Multi Layer Perceptron represented as a graph. There are three layers, respectively, input with black nodes, hidden with white nodes and the output represented by a single gray node. At each layer, its nodes are fully connected with each node of the next layer. Each connection is identified by a numerical value called \textit{weight}, usually a real number normalized in the range $[-1, +1]$.}\label{PHAT:mlp}
   \end{figure*}

A Multi-Layer Perceptron may be also represented by a graph, as also shown in Fig \ref{PHAT:mlp}: the input
layer ($x_i$) is made of a number of perceptrons equal to the number
of input variables ($d$); the output layer, on the other hand, will
have as many neurons as the output variables ($K$).  The network may
have an arbitrary number of hidden layers (in most cases one) which in
turn may have an arbitrary number of perceptrons ($M$).  In a fully
connected feed-forward network each node of a layer is connected to
all the nodes in the adjacent layers.

Each connection is represented
by an adaptive weight which represents the strength of the synaptic
connection between neurons ($w_{kj}^{(l)}$). The response of each
perceptron to the inputs is represented by a non-linear function $g$,
referred to as the activation function.  Notice that the above
equation assumes a linear activation function for neurons in the
output layer. We shall refer to the topology of an MLP and to the
weights matrix of its connections as to the model.  In order to find
the model that best fits the data, one has to provide the network with
a set of examples: the training phase thus requires the KB, i.e. the
training set.  The learning rule of our MLP is the Quasi Newton
Algorithm (QNA) which differs from the Newton Algorithm in terms of the calculation of the hessian of the error function.  In fact Newtonian models are
variable metric methods used to find local maxima and minima of
functions \citep{davidon1968} and, in the case of MLPs they can be
used to find the stationary (i.e. the zero gradient) point of the
learning function and are the general basis for a whole
family of so called Quasi Newton methods.

The traditional Newton method uses the Hessian of a function to find the
stationary point of a quadratic form. The Hessian of a function is not always available and in many cases it
is far too complex to be computed.  More often we can only calculate
the function gradient which can be used to derive the Hessian
via N consequent gradient calculations.

The gradient in every point w is in fact given by:

\begin{equation}
\nabla \mathbf{E} = \mathbf{H} \times \left( \mathbf{w}-\mathbf{w}^*\right)
\label{PHAT:eq2}
\end{equation}

where $\mathbf{w}^∗$ corresponds to the minimum of the error function, which satisfies the condition:

\begin{equation}
\mathbf{w}^*∗ = \mathbf{w} −- \mathbf{H}^{−-1} \times \nabla E
\label{PHAT:eq3}
\end{equation}

The vector $−\mathbf{-H}^{−-1} \times \nabla \mathbf{E}$ is known as
Newton direction and it is the traditional base for a variety of optimization
strategies,

Thus the step of this traditional method is defined as the product of an inverse Hessian matrix and a function
gradient. If the function is a positive definite quadratic form, the
minimum can be reached in just one step, while in case of an
indefinite quadratic form (which has no minimum), we will reach either
the maximum or a saddle point. To solve this problem, Quasi Newton
methods proceed with a positive definite Hessian approximation.  So
far, if the Hessian is positive definite, we make the step using the
Newton method.  If, instead it is indefinite, we first modify it to
make it positive definite, and then perform a step using the Newton
method, which is always calculated in the direction of the function
decrement.

In practice, QNA is an optimization of learning rule based on a
statistical approximation of the Hessian by cyclic gradient
calculation which, as already mentioned, is at the base of the classical Back
Propagation (BP; \citealt{bishop2006}) method.

The QNA instead of calculating
the H matrix and then its inverse, uses a series of intermediate steps
of lower computational cost to generate a sequence of matrices which
are more and more accurate approximations of $\mathbf{H}^{−-1}$.
During the exploration of the parameter space, in order to find the
minimum error direction, QNA starts in the wrong direction. This
direction is chosen because at the first step the method has to follow
the error gradient and so it takes the direction of steepest
descent. However, in subsequent steps, it incorporates information
from the gradient.  By using the second derivatives, QNA is able to
avoid local minima and to follow more precisely the error function
trend, revealing a "natural" capability to find the absolute minimum
error of the optimization problem.

However, this last feature could be a downside of the model,
especially when the signal-to-noise ratio of data is very poor. But
with "clean" data, such as in presence of high quality spectroscopic
redshifts, used for model training, the QNA performances result
extremely precise.

The experiment described in section 4 consists of a supervised regression based on
the MLP neural network trained by the Quasi Newton learning rule. As already described, the MLP is a network model composed by input and two computational layers of neurons (see Eq.~\ref{PHAT:eq1}), which propagate submitted data from input to output
layer. Each neuron of hidden layer is represented by a non-linear activation function
(in our case hyperbolic tangent) of the sum of inputs from all
previous layer neurons, multiplied by weights (normalized values in
[-1, +1] representing the connections between neurons, see
fig.~\ref{PHAT:mlp}).
After propagating the input data, at the final (output) layer, the
learning error is evaluated (in our case by means of the Mean Square
Error, MSE, between calculated vs desired outputs), and then the
backward phase is started, in which a learning rule is applied, by
adapting the neuron connection weights in such a way that the error
function is minimized. Then the input data are submitted again and a
new cycle of learning is achieved. The algorithm stops after a chosen
number of iterations or if the error becomes less than a chosen
threshold.  Note that the error is calculated at each iteration by
comparing the calculated value (on all input data) against the desired
(\textit{a priori} known) target value. This is the typical approach
called ``supervised''.  When the learning phase is stopped, the
trained network is used like a simple function. Input data not used
for training, or a mix in case of learning validation, can be
submitted to the network, that, if trained well, is able to provide
correct output (generalization capability).  By looking at the local
squared approximation of the error function, it is possible to obtain
an expression of minimum position. It is in fact known that the
gradient in every point w of the error surface is given by
Eq.~\ref{PHAT:eq2}.
The network is trained in order to learn to calculate the correct
photometric redshift given the input features for each object (see section 4).
This is indeed a typical supervised regression problem.\\
In terms of computational cost, the implementation of QNA can be problematic.
In fact to approximate the inverse Hessian matrix it requires to generate and to store $N\times N$ approximations, where $N$ is the number of variables and so the number of gradients involved in the calculation. So far, given $nI$ the number of iterations chosen by the user, the total computational cost is about $nI*N^2$ floating point per second (flops).
For this reason it exists a family of quasi-newton optimization methods, which allow to improve the complexity of the algorithm. In particular, in our implementation, we use the limited-memory BFGS (L-BFGS; \citealt{byrd1994,
  broyden1970, fletcher1970, goldfarb1970, shanno1970}), where BFGS is the acronym composed by the names of the four inventors.\\
L-BFGS never stores the full $N$ approximations of the hessian matrix, but only the last $M$ steps (with $M<<N$). Hence, given $M$ the stored approximation steps, the computational cost could be reduced to about $nI*(N*M)$ flops, which in practice trasforms the total cost of the algorithm from an exponential form to a polynomial one.
Moreover, in order to give a complete computational complexity evaluation for the implementation of the MLPQNA model, it remains to analyze the feed-forward part of the algorithm, for instance the computational flow of input patterns throughout the MLP network, up to the calculation of the network error (as said the MSE between the desired spectroscopic redshift and the one calculated by the network), at each training iteration after a complete submission of all input patterns.\\
The feed-forward phase involves the flow of each input pattern throughout the network, from the input to output layer, passing through the hidden layer.
This phase can be described by the following processing steps (\citealt{mizutani2001}):
\begin{itemize}
\item {\it Process 1 (P1)}: network node input computation;
\item {\it Process 2 (P2)}: network node activation function computation;
\item {\it Process 3 (P3)}: error evaluation;
\end{itemize}
The computational cost, in terms of needed flops, for the above three processing steps, can be summarized as follows.\\
Given $d$ the number of training data, $N_w$ the number of network weights, $A_f$ and $N_n$ respectively, the flops needed to execute the activation function (strongly depending on the hosting computer capabilities) and number of nodes present in the hidden plus output layers, $O_n$ the number of output nodes, we obtain:
\begin{equation}
P1 \cong d \times N_w
\label{PHAT:eq4}
\end{equation}
\begin{equation}
P2 \cong d \times A_f \times N_n
\label{PHAT:eq5}
\end{equation}
\begin{equation}
P3 \cong d \times O_n
\label{PHAT:eq6}
\end{equation}

In conclusion, the computational cost for the feed-forward phase of the MLPQNA algorithm has a polynomial form of about $nI*d \times [N_w+(A_f \times N_n)+O_n]$. The total complexity of MLPQNA implementation is hence obtained by the polynomial expansion of Eq.~\ref{PHAT:eq7}, as the sum of feed-forward and backward phases multiplied by the number of training iterations.
\begin{equation}
flops \cong nI*[(d \times (N_w+(A_f \times N_n)+O_n))+(N*M)]
\label{PHAT:eq7}
\end{equation}
Considering our training experiment described in Sect.~4.3 and using parameters reported in Tab.~\ref{Tab:exp1}, from Eq.~\ref{PHAT:eq7} we obtain about $1200$ Gflops, which corresponds to about $15$ min of execution time.

\section{The experiment Workflow}
\label{experiments}
In this section we describe the details of the sequence of concatenated computational steps performed in order to determine photometric redshifts. This is what we intended as a workflow, whick can be seen also as the description of the procedure building blocks.\\
MLPQNA method was applied by following the standard Machine Learning (ML) workflow (\citealt{bishop2006}),
which is here summarized: {\it i)}
extraction of the KB by using the 515 available spectroscopic
redshifts; {\it ii)} determination of the "optimal" model parameter
setup, including pruning of data features and training/test with the
available KB; {\it iii)} application of the tuned model to measure
photometric redshifts on the whole PHAT1 dataset of N=1984 objects, by
including also the re-training on the extended KB.  We also follow the
rules of the PHAT1 contest, applying the new method in two different
ways, first to the whole set of 18 bands and then to the 14 non-IRAC
bands only.
In order to better clarify what is deeply discussed in the next sub-sections, it is important to stress that the 515 objects, with spectroscopic redshifts publicly available, have been used to tune our model. In practice, 400 objects have been used as training set and the remaining 115 as test/validation set (steps {\it i)} and {\it ii)} of the workflow, see Sect.s~4.1,~4.2).
After having tuned our model, we performed a full training on all 515 objects, in order to exploit all the available knowledge base (see Sect.~4.3).

\subsection{Extraction of the Knowledge Base}
\label{sub:i}

For supervised methods it is common praxis to split the KB in at least
three disjoint subsets: one (training set) to be used for training
purposes, i.e. to teach the method how to perform the regression; the
second one (validation set) to check against loss of generalization
capabilities (also known as overfitting); and the third one (test set)
to be used to evaluate the performances of the model.  As a rule of
thumb, these sets should be populated with 60\%, 20\% and 20\% of the
objects in the KB, respectively. In order
to ensure a proper coverage of the Parameter Space (PS), objects in
the KB are split among the three datasets by random extraction and
usually this process is iterated several times in order to minimize
biases introduced by fluctuations in the coverage of the PS.

In the case of MLPQNA described here, we used cross-validation
(cf. \citealt{geisser1975}) in order to minimize the size of the
validation set ($\sim 10\%$). Training and validation were therefore
performed together using as training set $\sim 80\%$ of the objects
and as test set the remaining $\sim 20\%$ (in practice 400 records in
the training set and 115 in the test set).  In order to ensure a
proper coverage of the PS we checked that the randomly extracted
populations had a spec-z distribution compatible with that of the
whole KB.  The automatized process of the cross-validation was done by
performing 10 different training runs with the following procedure: (i) we split the training set into 10
random subsets, each one composed by 10\% of the dataset; (ii) at each training run we
apply the 90\% of the dataset for training and the excluded 10\% for validation.
This procedure is able to avoid overfitting on the training set (\citealt{bishop2006}).
There are several variants of cross validation methods \citep{sylvain2010}. We in particular have chosen the k-fold cross validation, particularly suited in presence of a scarcity of known data samples \citep{geisser1975}. Since the Eq.~\ref{PHAT:eq7} is referred to a single training run, in case of application of the k-fold cross validation procedure, the execution time could be estimated by multiplying the Eq.~\ref{PHAT:eq7} by the factor $k-1$, where $k$ is the total number of runs.

\subsection{Model optimization}
\label{sub:ii}

As known, supervised machine learning models are powerful methods able
to learn from training data the hidden correlation between input and
output features.  Of course, their generalization and prediction
capabilities strongly depend by the intrinsic quality of data
(signal-to-noise ratio), level of correlation inside of the PS and by
the amount of missing data present in the dataset.  Among the factors
which affect performances, the most relevant is the fact that most ML
methods are strongly sensitive to the presence of Not a Number (NaN)
in the dataset to be analysed (\citealt{vashist2012}).  This is especially relevant in
astronomical dataset where NaN's may either be non detections
(i.e. objects which in a given band are observed but non detected
since they are below the detection threshold) or related to patches of
the sky which have not been observed.  The presence of features with a
large fraction of NaN's can seriously affect the performances of a
given model and lower the accuracy or the generalization capabilities
of a specific model.  It is therefore a good praxis to analyze the
performances of a specific model in presence of features with large
fractions of NaN's. This procedure is strictly related to the so
called feature selection or ''pruning of the features'' phase which
consists in evaluating the significance of individual features to the
solution of a specific problem.  In what follows we shall shortly
discuss the outcome of the ''pruning'' performed on the PHAT1 dataset.

\subsubsection{Pruning of features}
\label{subsub:pru}

It is also necessary to underline that especially in presence of small datasets there is a need for a compromise: while on the one hand it is necessary to minimize the effects of NaN's, on the other it is not possible to simply remove each record containing a NaN, because otherwise too much information would be lost.

In table \ref{Tab:pruning} we list the percentage of NaN's in each photometric band both in the training and full datasets. Poor features, namely the fluxes in the K and m5.8 bands were not used for the subsequent analysis.

\begin{table*}
\begin{center}
\begin{tabular}{|c|c|c|c|c|c|c|}
  \hline
  % after \\: \hline or \cline{col1-col2} \cline{col3-col4} ...
BAND & Dataset Column ID & \% NaN in whole set & \% NaN in Training & NaN \% Absolute Difference
\\ \hline m5.8 & 17 & 19.35 & 17.28 & 2.07
\\K & 14 & 17.14 & 18.64 & 1.5
\\HK & 13 & 5.65 & 6.21 & 0.57
\\m8 & 18 & 3.48 & 3.5 & 0.02
\\F435W & 7 & 2.67 & 1.75 & 0.92
\\H & 12 & 2.37 & 2.52 & 0.16
\\J & 11 & 1.16 & 1.55 & 0.39
\\U & 1 & 1.01 & 1.17 & 0.16
\\R & 4 & 0.15 & 0.19 & 0.04
\\B & 2 & 0.1 & 0.19 & 0.09
\\V & 3 & 0.05 & 0.19 & 0.14
\\F606W & 8 & 0.05 & 0 & 0.05
\\m 3.6 & 15 & 0.05 & 0 & 0.05
\\I & 5 & 0 & 0 & 0
\\Z & 6 & 0 & 0 & 0
\\F775W & 9 & 0 & 0 & 0
\\F850LP & 10 & 0 & 0 & 0
\\m4.5 & 16 & 0 & 0 & 0
\\
  \hline
\end{tabular}\end{center}
  \caption{The percentages of Not a Number in the whole dataset (col 3), with 1984 objects and in the trainset (col 4), with 515 objects, for each band. The last column reports the absolute differences between the two NaN percentages. As shown this difference remains always under 3\%, demonstrating that the two datasets are congruent in terms of NaN quantity.}\label{Tab:pruning}

\end{table*}

The pruning was performed separately on the two PHAT1 datasets (18-bands and 14-bands), respectively. A total of 37 experiments was run on the two datasets: the various experiments differing in the groups of features removed. We started by considering all features (bands), removing the two worst bands, for instance K and m5.8, which outlier quantity was over the 15\% of patterns. Then a series of experiments was performed by removing one band at a time, by considering the NaN's percentage shown in table \ref{Tab:pruning}.

\subsubsection{Performance metrics}
\label{subsub:met}

The performances of the various experiments were evaluated (as done in the PHAT contest) in terms of:
\begin{itemize}
\item {\it scatter}: is the RMS of $\Delta z$
\item {\it bias}: is the mean of $\Delta z$
\item {\it fraction of outliers}: where outliers are defined by the condition: $\left|\Delta z \right| > 0.15$
\end{itemize}
Where:
\begin{equation}
\Delta z \equiv \frac{z_{spec} - z_{phot}}{1+z_{spec} }
\label{PHAT:eq8}
\end{equation}

\begin{table*}
\begin{center}
\begin{tabular}{|l|rrrrrrrrrrr|}
\hline
exp. n & missing features & feat. & hid.& step	&res. &	dec.   &	MxIt&	CV&	scatter&	outliers$\%$&	bias  \\
\hline
37	   & m5.8,K, HK, m8	 & 14	   & 29	  & 0.0001&	30	& 0.1	 & 3000	  & 10& 0.057	 & 22.61\%  & -0.0077\\
26	   & m5.8, K, m3.6, m4.5, HK, m8& 12 & 25	  & 0.0001&	30	& 0.1	 & 3000	  & 10&	0.062	 & 17.39\% 	& 0.0078 \\
\hline
\end{tabular}
\end{center}
\caption{Description of the best experiments for the  18 bands (Exp. n. 37) and the 14 bands datasets (Exp. n. 26).
Column 1: sequential experiment identification code; column 2: features not used in the experiment; columns 3-4: number
of input (features) and hidden neurons; column 5--9: parameters of the MLPQNA used during the experiment; column 10: scatter error evaluated as described in the text; column 11: fraction of outliers; column 12: bias.}\label{Tab:exp1}
\end{table*}

At the end of this process, we obtained the best results, reported in table \ref{Tab:exp1}.

\subsection{Application to the PHAT1 dataset}

We performed a series of experiments in order to fine tune the model
parameters, whose best values are:

MLP network topology parameters (see Tab.~\ref{Tab:exp1}):
\begin{itemize}
\item feat: 14 (12) input neurons (corresponding to the pruned number of input band magnitudes listed in Tab.~\ref{Tab:pruning});
\item hid: 29 (25) hidden neurons;
\item 1 output neuron.
\end{itemize}

QNA training rule parameters (see Tab.~\ref{Tab:exp1}):
\begin{itemize}
\item step: 0.0001 (one of the two stopping criteria. The algorithm stops if approximation error step size is less than this value. A step value equal to zero means to use the parameter MxIt as unique stopping criterion.);
\item res : 30 (number of restarts of hessian approximation from random positions, performed at each iteration);
\item dec : 0.1 (regularization factor for weight decay. The term $dec*||network weights||^2$ is added to the error function, where $network weights$ is the total number of weights in the network. When properly chosen, the generalization error of the network is highly improved);
\item MxIt: 3000 (max number of iterations of hessian approximation. If zero the step parameter is used as stopping criterion);
\item CV: 10 (k-fold Cross Validation, with k=10. This parameter is described in section 4.1).
\end{itemize}

With such parameters, we obtained the statistical results (in terms of scatter, bias and outlier percentage) as reported in the last three columns of Tab.~\ref{Tab:exp1}.

Once the model optimization described above had been
determined, the MLPQNA was re-trained on the
whole KB (515 objects) and applied to the whole PHAT1 dataset
(1984 objects), which was then submitted to the PHAT contest
for final evaluation (see below).

Details of the experiments can be found at the DAME web site\footnote{\url{http://dame.dsf.unina.it/dame_photoz.html}}, while the parameter settings and the results for the best models are summarised in table \ref{Tab:results}.

\section{The PHAT1 results and comparison with other models}\label{PR}

With the model trained as described in the above section, we calculated photometric redshifts for the entire PHAT1 dataset, i.e. also for the remaining 1469 objects, for which the corresponding spectroscopic redshift was hidden to the contest participants, obtaining a final photometric catalogue of 1984 objects. This output catalogue has been finally delivered to PHAT contest board, receiving as feedback the statistical results (scatter, bias and outlier's percentage) coming from the comparison between spectroscopic and photometric information, in both cases (18 and 14 bands).\\
So far, the statistical results and plots referred to the whole data sample, which is kept secret to all participants as required by the PHAT contest, were provided by H. Hildebrandt and reported also in the PHAT Contest wiki site \footnote{\url{http://www.astro.caltech.edu/twiki_phat/bin/view/Main/GoodsNorthResults#Cavuoti_Stefano_et_al_neural_net}}.
So far, the results obtained by analysing the photometric redshifts calculated by MLPQNA, are shown in table \ref{Tab:results}.

The most significative results can be summarized as it follows:

\begin{description}

\item[i)] 18-band experiment: 324 outliers with $\left|\Delta_z\right| > 0.15$, corresponding to a relative fraction of $16.33\%$.
For the remaining 1660 objects bias and rms are: $0.000604251 \pm 0.0562278$

\item[ii)] 14-band experiment: 384 outliers with $\left|\Delta_z\right| > 0.15$, corresponding to a relative fraction of $19.35\%$.
1600 objects with bias and variance $ 0.00277721 \pm  0.0626341$.

\end{description}

\begin{figure*}
   \centering
   \includegraphics[width=7cm]{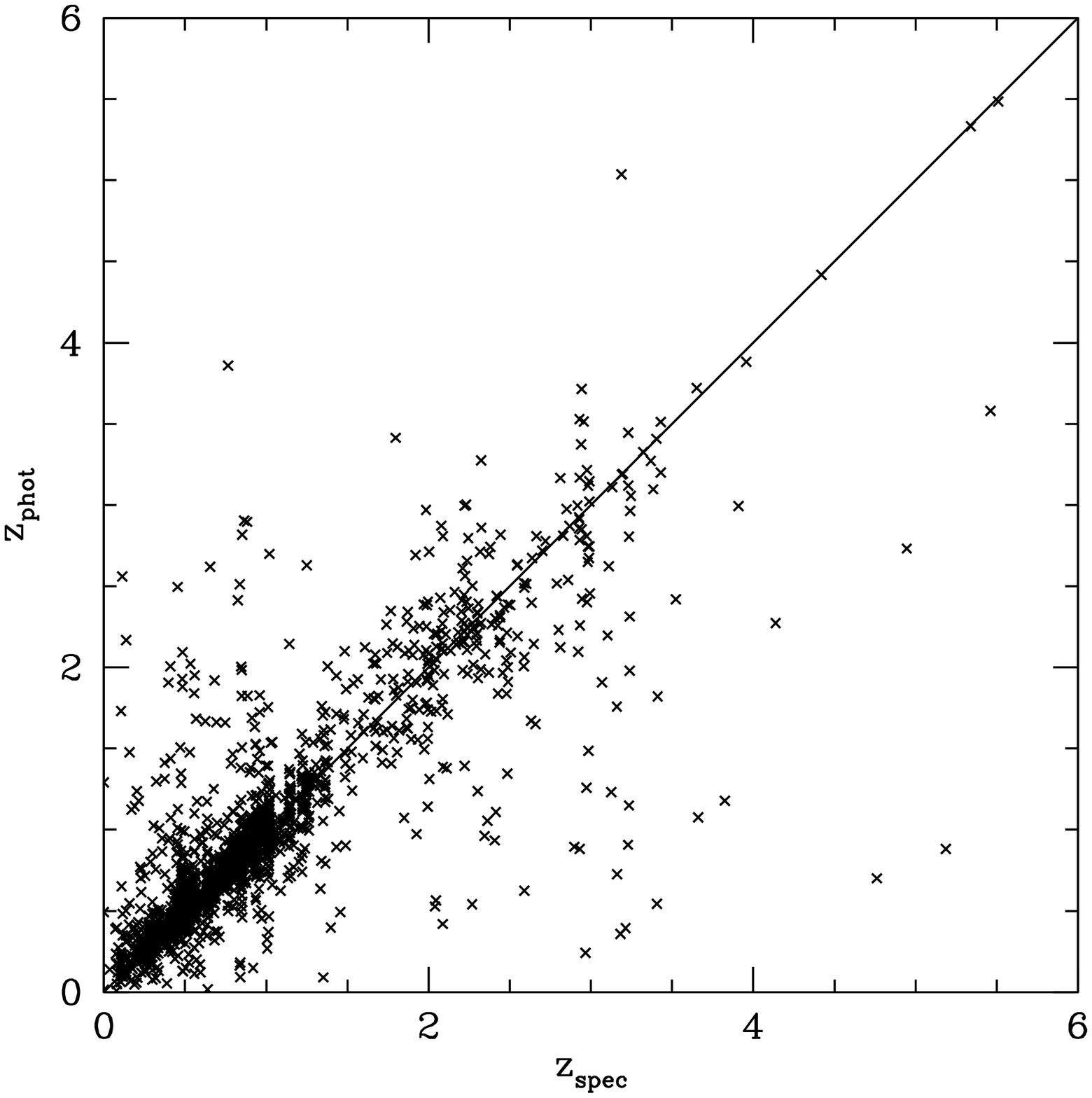} (a)
   \includegraphics[width=7cm]{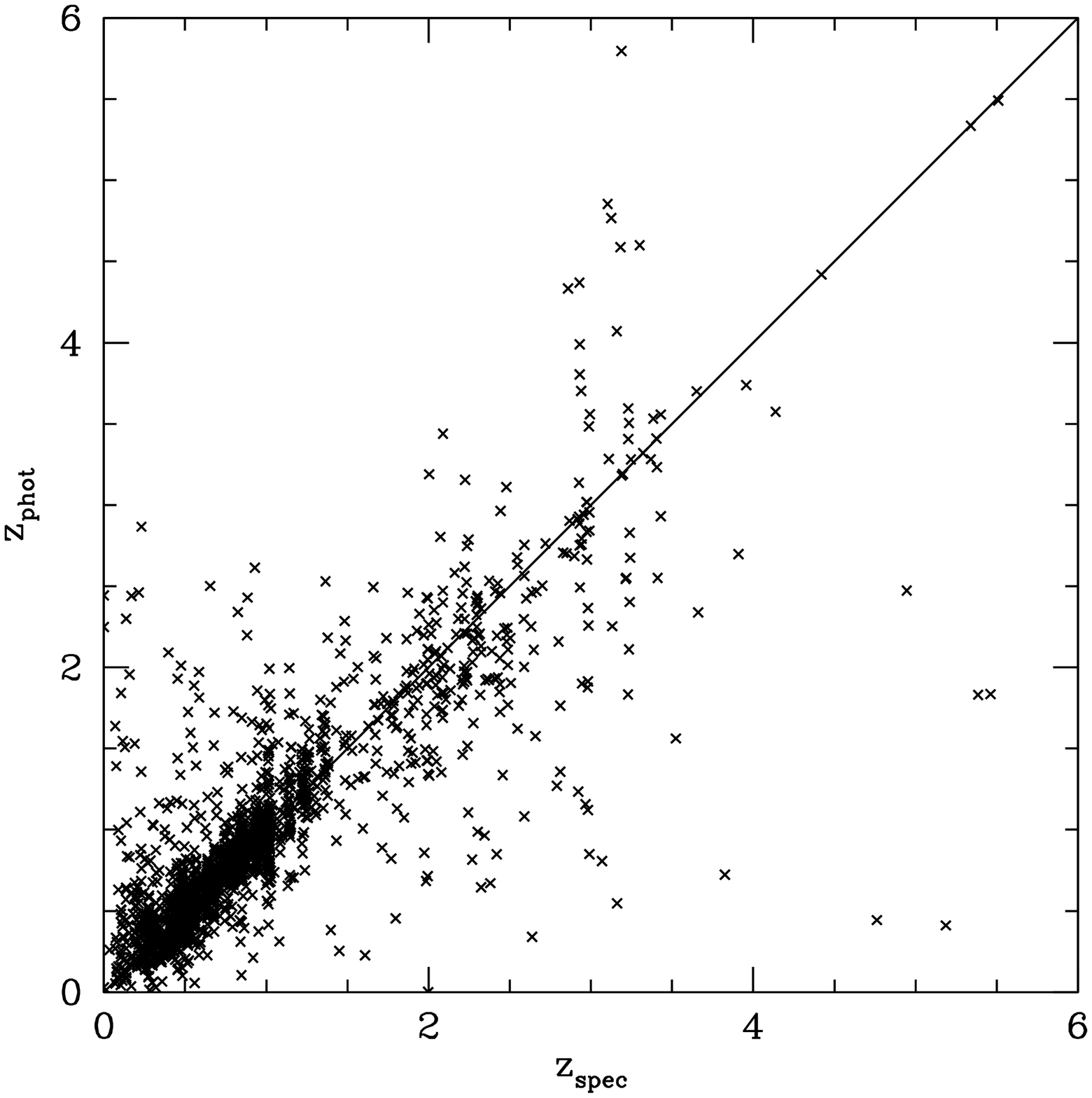} (b)
   \caption{Results obtained by our model and provided by the PHAT contest board in terms of direct comparison between our photometric and blind spectroscopic information.  In the (a) panel are plotted
     the photometric vs. spectroscopic redshifts for the whole dataset
     using 10 photometric bands (Experiment 37). In panel (b) the same
     but using only 14 photometric bands (Experiment
     26). (Courtesy of H. Hildebrandt).}\label{PHAT:scatter}
   \end{figure*}

A more detailed characterization of the results can be found in the
first line of parts A, B and C in the table \ref{Tab:results}, while
figure \ref{PHAT:scatter}, provided by H. Hildebrandt, gives the scatter plots (spec-z's vs
photo-z's) for the 18 and 14 bands, respectively.

\begin{table*}
\begin{center}
\begin{tabular}{|l|rrr|rrr|rrr|rrr|}
\hline
\hline
{\bf A}	    &\multicolumn{3}{|c|}{18-band; $\left|\Delta z\right| \leq 0.15$}
            &\multicolumn{3}{|c|}{14-band; $\left|\Delta z\right| \leq 0.15$}				
            &\multicolumn{3}{|c|}{18-band; R$< 24$; $\left|\Delta z\right| \leq0.15$}
            & \multicolumn{3}{|c|}{14-band; R$< 24$; $\left|\Delta z\right| \leq 0.15$}\\
\hline		
Code	& bias	  &scatter& outliers \% &	bias	&scatter &outliers \% &bias	  &scatter &outliers \% &bias	    &scatter &outliers \% \\
\hline
QNA	    & 0.0006 &0.056	&16.3	& 0.0028 &0.063	&19.3	& 0.0002 &0.053	&11.7	& 0.0016	&0.060	&13.7\\
AN-e	&-0.010	 &0.074	&31.0	&-0.006 &0.078	&38.5	&-0.013	&0.071	&24.4	&-0.007	&0.076	&32.8\\
EC-e	&-0.001	 &0.067	&18.4	& 0.002 &0.066	&16.7	&-0.006	&0.064	&14.5	&-0.003	&0.064	&13.5\\
PO-e	&-0.009	 &0.052	&18.0	&-0.007 &0.051	&13.7	&-0.009	&0.047	&10.7	&-0.008	&0.046	& 7.1\\
RT-e	&-0.009	 &0.066	&21.4	&-0.008 &0.067	&24.2	&-0.012	&0.063	&16.4	&-0.012	&0.064	&18.4\\
\hline
{\bf B}     &\multicolumn{3}{|c|}{18-band; $\left|\Delta z\right| \leq 0.5$}
            &\multicolumn{3}{|c|}{14-band; $\left|\Delta z\right| \leq 0.5$}
            &\multicolumn{3}{|c|}{18-band; R$< 24$; $\left|\Delta z\right| \leq 0.5$}
            &\multicolumn{3}{|c|}{14-band; R$< 24$; $\left|\Delta z\right| \leq 0.5$}\\
\hline		
Code	& bias	  &scatter & outliers \% &	bias	&scatter & outliers \% &bias	  &scatter &outliers \% &bias	    &scatter & outliers \% \\
\hline
QNA	  &-0.0028	&0.114	&3.8	&-0.0046 &0.125	&3.8	 &-0.0039  &0.101	&1.7	&-0.0039 &0.101	&1.7\\
AN-e	&-0.036	&0.151	&3.1	&-0.035	&0.173	&4.2	 &-0.047	&0.130	&1.4	&-0.047	&0.130	&1.4\\
EC-e	&-0.007	&0.120	&3.6	&-0.003	&0.114	&3.6	 &-0.015	&0.106	&1.9	&-0.015	&0.106	&1.9\\
PO-e	&-0.013	&0.124	&3.1	&0.001	&0.107	&2.3	 &-0.020	&0.098	&1.2	&-0.020	&0.098	&1.2\\
RT-e	&-0.031	&0.126	&3.2	&-0.028	&0.137	&3.6	 &-0.034	&0.111	&1.4	&-0.034	&0.111	&1.4\\
\hline
{\bf C}      &\multicolumn{3}{|c|}{18-band; $z_{sp}\leq 1.5$, $\left|\Delta z\right| \leq 0.15$}
             &\multicolumn{3}{|c|}{14-band; $z_{sp}\leq 1.5$,  $\left|\Delta z\right| \leq 0.15$}				
             &\multicolumn{3}{|c|}{18-band; $z_{sp}>1.5$,      $\left|\Delta z\right| \leq 0.15$}
             &\multicolumn{3}{|c|}{14-band; $z_{sp}>1.5$,      $\left|\Delta z\right| \leq 0.15$}\\
\hline
Code	& bias	  &scatter& outliers \% &	bias	 & scatter & outliers \% &	bias	&scatter &outliers \% &bias	  &scatter & outliers \% \\
\hline
QNA	  &-0.0004	&0.053	&14.6	&0.0001   &0.061	&16.6	&0.0074	&0.072	&26.3	&0.0222	&0.070	&35.0\\
AN-e	&-0.017  &0.070  &27.6	&-0.010	  &0.076	&33.6	&0.051	&0.078	&50.7	&0.045	&0.077	&66.4\\
EC-e	&-0.003	&0.065  &16.1	&-0.000	  &0.064	&14.5	&0.015	&0.077	&32.3	&0.015	&0.077	&29.5\\
PO-e	&-0.012	&0.049  &12.6	&-0.011	  &0.047	&9.4	&0.019	&0.075	&48.3	&0.026	&0.074	&37.7\\
RT-e	&-0.016	&0.062	&19.6	&-0.014	  &0.064	&21.1	&0.040	&0.072	&31.8	&0.039	&0.071	&41.9\\
\hline
\end{tabular}
\end{center}
\caption{Comparison of the performances of our MLPQNA (here labeled as QNA) method against all other empirical methods analysed by PHAT board.
For a description of other methods (namely AN-e, EC-e, PO-e and RT-e) see the text. The table is divided into three parts
(namely A, B and C). Data for the other empirical method have been extracted from \cite{hildebrandt2010}.
In each part of the table we list the results (on both the 18 and the 14 bands datasets) for a specific subsample of the PHAT objects.
Part A: statistical indicators (bias and scatter) for the 18 and 14 bands computed on objects with $\left|\Delta z\right| \leq 0.15$ and for objects with $\left|\Delta z\right|\leq 0.15$ and $R<24$. The column ``outliers'' gives the fraction of outliers defined as objects with $\left|\Delta z\right| >0.15$.
Part B: the same but for $\left|\Delta z\right| \leq 0.5$.
Part C: the same but for objects with spectroscopic redshift $z_{sp} \leq 1.5$ and $\left|\Delta z\right| \leq 1.5$, and for $z_{sp} > 1.5$ and $\left|\Delta z\right| \leq 1.5$. The definitions of bias, scatter and outlayers fraction are given in the text.
Values were computed by the PHAT collaboration on the whole PHAT1 dataset.}\label{Tab:results}
\end{table*}

In order to compare our results with other models, we also report in
table \ref{Tab:results} the statistical indicators for the other
empirical methods which competed in the PHAT1 contest.  The methods
are:

\begin{itemize}
\item {\bf AN-e}: ANNz, Artificial Neural Network, an empirical photo-z code
  based on artificial neural networks  \citep{collister2004};
\item {\bf EC-e}: Empirical $\chi^2$, a subclass of kernel regression
  methods; which mimics a template-based technique with the main
  difference that an empirical dataset is used in place of the
  template grid  \citep{wolf2009};
\item {\bf PO-e}: Polynomial Fit, a ''nearest neighbour'' empirical
  photo-z method based on a polynomial fit so that the galaxy redshift
  is expressed as the sum of its magnitudes and colours \citep{li2008};
\item {\bf RT-e}: Regression Trees, based on Random Forests which
  are an empirical, non-parametric regression technique \citep{carliles2010}.
\end{itemize}

More details can be found in the quoted references and in
\cite{hildebrandt2010}.

For each of the datasets (18 and 14 bands), statistics in Table
\ref{Tab:results} refers to several regimes: the first one (A) defines
as outliers all objects having $\left| \Delta z \right| > 0.15$ and it
is divided into two subsections: the left side includes all objects,
while the right side includes objects brighter than R = 24; the second
one (B) defines as outliers objects having $\left| \Delta z \right| >
0.50$ and it is divided as section (A); the third one (C) defines as
outliers objects having $\left| \Delta z \right| > 0.50$ and divided
into a left side, for object with $z \leq 1.5$ and a right side having
$z > 1.5$.

By analyzing the MLPQNA performance in the different regimes, we obtained:

{\it All objects}: in the 18 bands experiment, QNA scores the
  best results in term of bias, and gives comparable results with PO-e
  in terms of scatter and number of outliers. In fact, while in Part A
  the scatter is slightly larger than those of PO-e method (0.052
  against 0.056), the number of outliers is lower (18.0\% against
  16.3\%) and in Part. B is the viceversa (0.124 against 0.114 and
  3.1\% against 3.8\%). In the 14 band experiment QNA obtains values
  slightly higher than PO-e in terms of scatter (0.051 against 0.063)
  and than EC-e in terms of bias (0.002 against 0.0028). For what concerns
  the fraction of outliers QNA scores results larger than PO-e
  and EC-e (13.7\% and 16.7\% against 19.3\%).

{\it Bright objects}: for brigth objects (R$<$24), the QNA resulting bias is again the best within the different empirical methods,
  while for scatter and number of outliers, QNA obtains values slightly
  higher than PO-e in both the 18 (0.047 against 0.053 and 10.7\%
  against 11.7\%) and the 14 bands datasets (0.046 against 0.060 and
  7.1\% against 13.7\%).

{\it Distant vs near objects}: in the distant sample
  ($z_{sp}>1.5$) QNA scores as first in terms of bias, scatter, and
  number of outliers for 18 bands. In the 14 band dataset case, it results the best method in
  terms of scatter, but with a bias (0.015 against 0.0222)
  and number of outliers (29.5\% against 35.0\%) higher than EC-e.
  In the near sample ($z_{sp}<1.5$) QNA is the best in terms of bias. The scatter is slightly higher than PO-e's for both 18 (0.049 against 0.053) and 14 bands (0.047
  against 0.061). For what concerns outliers, PO-e performs better at
  18 bands (12.6\% against 14.6\%), while PO-e and EC-e perform
  better at 14 bands (9.4\% and 14.5\% against 16.6\%).

\section{Summary and Conclusions}
\label{discus}
For the first time the MultiLayer Perceptron with Quasi Newton learning rule described here has been exploited to solve regression problems in the astrophysical context.
This method was applied on the whole PHAT1 dataset of N=1984 objects \cite{hildebrandt2010} ) to determine photometric redshifts after an optimization of the model performed by using as a training set the 515 available spectroscopic redshifts.

The statistics obtained by the PHAT board, by analyzing the photometric redshifts derived with MLPQNA, and the comparison with other
empirical models are reported in Table \ref{Tab:results}.

From a quick inspection of table \ref{Tab:results}, it descends that it does not
exist an empirical method which can be regarded as the best in terms of all the indicators (e.g.
bias, scatter and number of outliers) and that EC-e (Empirical $\chi^2$ method), PO-e (Polynomial Fit method) and MLPQNA produce comparable results.
However, the MLPQNA method, on average, gives the best result in terms of bias at any regime.

For what the scatter is concerned, by considering the dataset with 18 bands reported in Parts A and B of  table \ref{Tab:results}, MLPQNA obtains
results comparable with the PO-e method.
In fact, in Part A PO-e's scatter is better than MLPQNA's, but with a larger number of outliers; while the trend is reversed in Part B.
In the other cases both the scatter and number of outliers are slightly worse than PO-e and EC-e
methods.

In general, MLPQNA seems to have better generalization capabilities
than most other empirical methods especially in presence of
underpopulated regions of the Knowledge Base.
In fact, $\sim 500$ objects with spectroscopic redshifts spread over such a large redshift interval are by far not sufficient to train most other empirical codes on the data.
This was also  pointed out also by \cite{hildebrandt2010}, who noticed that the high fraction of
outliers produced by empirical methods is on average higher than what is
currently found in literature ($\sim 7.5 \%$) and explained it
as an effect of the small size of the training sample, which maps
poorly the very large range in redshifts and does not include a large
enough number of objects with peculiar SED's.

In this respect we wish to stress that as it has already been shown in
another application (cf. \citealt{brescia2012}) and will be more extensively discussed in a forthcoming paper,
MLPQNA enjoys the very rare prerogative of being able to obtain good performances also when
the KB is small and thus undersampled (Brescia et al. in preparation).

\noindent {\it Acknowledgements.}\\
The authors wish to thank the anonymous referee for useful comments which improved the paper.
The authors wish also to thank H. Hildebrandt for the courtesy to provide statistical results and plots,
all present and former members of the DAME collaboration and in particular R. D'Abrusco for useful
discussions. DAME has been funded through several grants which are
mentioned on the collaboration web site. We acknowledge the use of
TOPCAT and of other Virtual Observatory tools during some steps of the
procedure. GL wishes to thank the California Institute of Technology
for the kind hospitality. AM and MB wish to thank the financial
support of PRIN-INAF 2010, "Architecture and Tomography of Galaxy Clusters".

\end{document}